# A systematic investigation of molecular encoding methods for drug property predictions across neural network and Transformer encoder-based model

**Sheng-Ya Chen[34] and Shan-Ju Yeh[123*]**

[1]School of Medicine, National Tsing Hua University, Hsinchu, Taiwan.
[2]Institute of Bioinformatics and Structural Biology, National Tsing Hua University, Hsinchu, Taiwan.
[3]Department of Life Science, National Tsing Hua University, Hsinchu, Taiwan.
[4]Interdisciplinary Program of Life Sciences and Medicine, National Tsing Hua University, Hsinchu, Taiwan.

[*]Corresponding author
E-mail: sjyeh@life.nthu.edu.tw (Shan-Ju Yeh)

## Abstract

Fundamental investigations into how different molecular encoding methods affect molecular property prediction remain relatively limited. In this study, we extensively examined the optimal molecular encoding methods for molecular properties prediction using two prevalent structure designs: a classical neural network model (MLP) and a Transformer encoder-based model (MLP+TL). For molecular encoding methods, we investigated several types of fingerprints, including traditional topological fingerprints, substructure-based fingerprints, and string-based representations. These two models were trained on seven well-known molecular datasets to evaluate different input molecular encoding methods based on evaluation metrics. On several biologically relevant classification tasks, including toxicity, mutagenicity, and side-effect prediction, our models consistently achieved average AUC values above 0.9. Beyond model performance, we further explored whether the Transformer encoder-based framework could provide interpretable insight. Rather than relying on external post-hoc explanation methods such as the local interpretable model-agnostic explanation (LIME) or the Deep SHapley Additive exPlanations (SHAP), we leveraged the model's intrinsic attention weights as an internal interpretability signal for identifying potentially important feature. To further support this analysis, we conducted case studies on the BBBP and MUTAG datasets. The MLP+TL model using MACCS and PubChem as input can capture chemically interpretable groups that determined the major blood-brain barrier (BBB) permeability and mutagenicity in *Salmonella typhimurium*. In particular, a comparison between Morphine and Heroin highlighted the role of hydroxyl-related substructures in BBB permeability prediction, which was consistently reflected in the attention weights. Overall, our findings provide practical guidance for selecting effective molecular encoding methods and contribute to the development of interpretable molecular informatics approaches for drug discovery.

# Introduction

Drug discovery [1, 2] is a lengthy, costly, and high-risk process, with a substantial attrition rate throughout the development pipeline. As a result, there has been growing interest in leveraging artificial intelligence (AI) and machine learning (ML) to improve the efficiency of early-stage drug discovery. By enabling the *in silico* evaluation of candidate compounds, predictive models can help eliminate unfavorable molecules at an earlier stage, prioritize more promising candidates, and thereby reduce the costs associated with blind synthesis, in vitro assays, animal studies, and late-stage failure. In this context, accurate molecular property prediction has become a crucial task in a wide range of applications, including drug discovery, toxicity assessment [3], and drug–drug interaction prediction [4]. A key prerequisite for such prediction tasks is the selection of an appropriate molecular representation, because the performance of downstream models strongly depends on how molecular structures are encoded into machine-readable forms. More generally, molecular representation can be understood as any encoding of a chemical compound, and different representations preserve different aspects of chemical information, which in turn affects model behavior and predictive performance. In recent years, substantial effort has been devoted to molecular representation learning, deep learning, and pretrained models to improve predictive performance [5]. However, recent studies have pointed out that benchmark-driven performance gains do not always reflect meaningful advances in practical applications [6]. In particular, prior work [7] has emphasized that model evaluation in molecular property prediction can be strongly influenced by factors such as statistical rigor, model applicability, evaluation metric selection, dataset split variability, and the extent to which external or chemical-space generalization is properly assessed. Moreover, improved metrics may sometimes reflect statistical fluctuation rather than robust methodological progress. Therefore, model development should not be guided solely by predictive performance, but should also consider interpretability, robustness, and generalizability. At the same time, although simpler machine learning models based on fixed molecular representations often remain competitive, they may be limited in their ability to capture richer structural patterns [8]. Motivated by this perspective, we focused on two neural network-based architectures (Figure 1) that have shown broad utility across domains: a multi-layer perceptron (MLP) [9] and a Transformer encoder-based model (MLP+TL) [10]. These two models were trained on seven well-known molecular datasets to evaluate different molecular encoding methods based on model performance. Rather than relying exclusively on external post hoc explanation methods, such as LIME[11] or SHAP[12, 13], we further explored the model's internal attention weights as an interpretability-related signal for identifying potentially important input features. In this way, the present study aims not only to evaluate predictive performance across molecular representations, but also to investigate whether an intermediate modeling strategy can provide a practical balance between predictive capability and interpretability.

From a representation perspective, molecules can be organized into three broad categories: fixed representation, which abstract structural patterns into predefined or hash-generated features; linear notations, exemplified by SMILES; and graph representations, which preserve the connectivity between atoms and bonds in an explicit graph form [1]. Among the various molecular representations available, molecular fingerprints are one of the most widely used fixed representations in cheminformatics, because they encode molecular structures into vectors that can be directly used in similarity analysis and machine learning models. Based on the kinds of chemical information that each type of fingerprint captures, five categories were taken into consideration [14]. By examining the paths taken by a pair of atoms across the molecular graph and hashing them inside a fixed-size vector, path-based fingerprints provide molecular features [15]. For instance, Depth First Search (DFS) [16] starts at each atom and explores up to a certain number of edges, representing a compound by capturing all unique paths in its molecular graph. Atom Pair fingerprints (AP) [17], which describe molecules by gathering all potential triplets of two atoms and the shortest path connecting them, are another example of this type of algorithm. Furthermore, a subset of path-based fingerprinting known as pharmacopohore fingerprints defines atoms depending on whether or not they are pharmacophore points (e.g., donors or acceptors of hydrogen bonds) [18]. This results in bit vectors that attempt to encode the molecule's interactions with its chemical surroundings rather than being as closely tied to the complex structure. Substructure-based fingerprints assign each bit to a predefined structural motif, indicating whether that motif is present in a molecule. [19, 20]. The PubChem fingerprints and the MACCS keys [19] are two examples of this class of algorithms. Like substructure-based fingerprints, circular fingerprints dissect a target compound into many fragments; however, rather than depending on expert-defined structural patterns, they dynamically generate them from the molecular graph of each molecule [21, 22]. To accomplish this, they first represent every atom in terms of certain characteristics, including atomic mass or valence. The numerical identifiers of nearby atoms are then appended to each atom, creating a fragment identifier. This procedure can be carried out multiple times, gradually consider expanding the neighborhood's radius when compiling data. Ultimately, a fixed-size vector is created by hashing every unique fragment for a specific molecule. In general, the distinction between fingerprints in this class can be seen in the use of distinct atom identifier attributes. Representative examples include extended-connectivity fingerprints (ECFP) [21] and functional-connectivity fingerprints (FCFPs) [21]. By working on the compound's simplified molecular input line entry system (SMILES) string rather than its graph representation, string-based fingerprints provide molecular representations [23, 24]. For example, LINGO fingerprints split SMILES strings into fixed-length substrings and encode each molecule according to the presence or frequency of these substrings. [24]. Next, using either counts or binary values, each compound is encoded based on which SMILES substrings in the set it contains. With the advent of advanced machine learning techniques, new encoding methods are continually being developed. Since functional groups in molecules play a pivotal role in determining their chemical properties and biological activities, selecting an appropriate molecular representation that captures these nuances is essential for accurate predictions. Despite the availability of numerous encoding methods, it remains unclear which representation yields the most accurate predictions across different applications. This knowledge gap motivates our investigation into identifying the optimal molecular encoding methods.

Multi-layer Perceptron (MLP) is a class of feedforward [9] artificial neural network that consists of at least three kinds of layers: an input layer, one or more hidden layers [25], and an output layer. Each layer is fully connected to the next one, and each node (or neuron) uses a nonlinear activation function [26], making MLPs capable of modeling complex relationships in data. MLPs have been foundational in the development of neural network research and are widely used in various applications, such as image recognition [27], and speech recognition [28]. Self-attention mechanism was first introduced in the context of the Transformer architecture [10]. This mechanism allows the model to weigh the importance of different words in a sentence relative to each other, regardless of their distance apart. This innovation has revolutionized natural language processing (NLP) [25, 29] by enabling models to capture long-range dependencies more effectively than previous methods. The concept of self-attention rapidly gained popularity following the introduction of the Transformer model. Notably, bidirectional encoder representations from transformers (BERT) [30] and generative pre-trained Transformer (GPT) [31] are some of the most influential models based on the Transformer architecture, leading to state-of-the-art performance on a wide range of NLP tasks. BERT, introduced by Devlin et al., demonstrated the power of bidirectional training of Transformers on masked language modeling, while GPT, developed by OpenAI, showed the effectiveness of generative pre-training on large text corpora. In addition, Transformers have been adapted for various applications, including image processing (e.g., Vision Transformers or ViTs) [32], drug discovery, and protein structure prediction [33]. For instance, the AlphaFold model by DeepMind, which uses Transformer-like architectures, has made significant advancements in predicting protein folding.

Based on the pipeline (Figures 2-3), our study produced several major findings. The first is a comparative analysis of molecular property prediction using MLP and MLP+TL, which reveals how model choice influences the effectiveness of different molecular encoding methods. The second is external validation for BBB penetration and solubility prediction, which evaluates whether the learned models remain robust when transferred to independent datasets beyond the internal splits. The third is dataset-level identification of representative substructures through attention-score analysis, where attention weights are used to highlight molecular fingerprint regions that contribute strongly to prediction. The fourth is a case comparison of Morphine and Heroin based on attention score-derived substructure patterns, which provides a more chemically intuitive illustration of how the model distinguishes closely related compounds. Overall, the goal of this study is to provide a fundamental investigation into which molecular encoding methods most effectively capture chemically informative patterns and yield strong predictive performance across different molecular property prediction tasks. In addition, by integrating Transformer-encoder layer with MLP model, the proposed framework seeks to preserve model simplicity while improving representational power and interpretability. From this perspective, the learned attention weights offer a direct way to examine how the model distributes focus across fingerprint features, thereby helping identify potentially important substructures interactions underlying the final prediction.

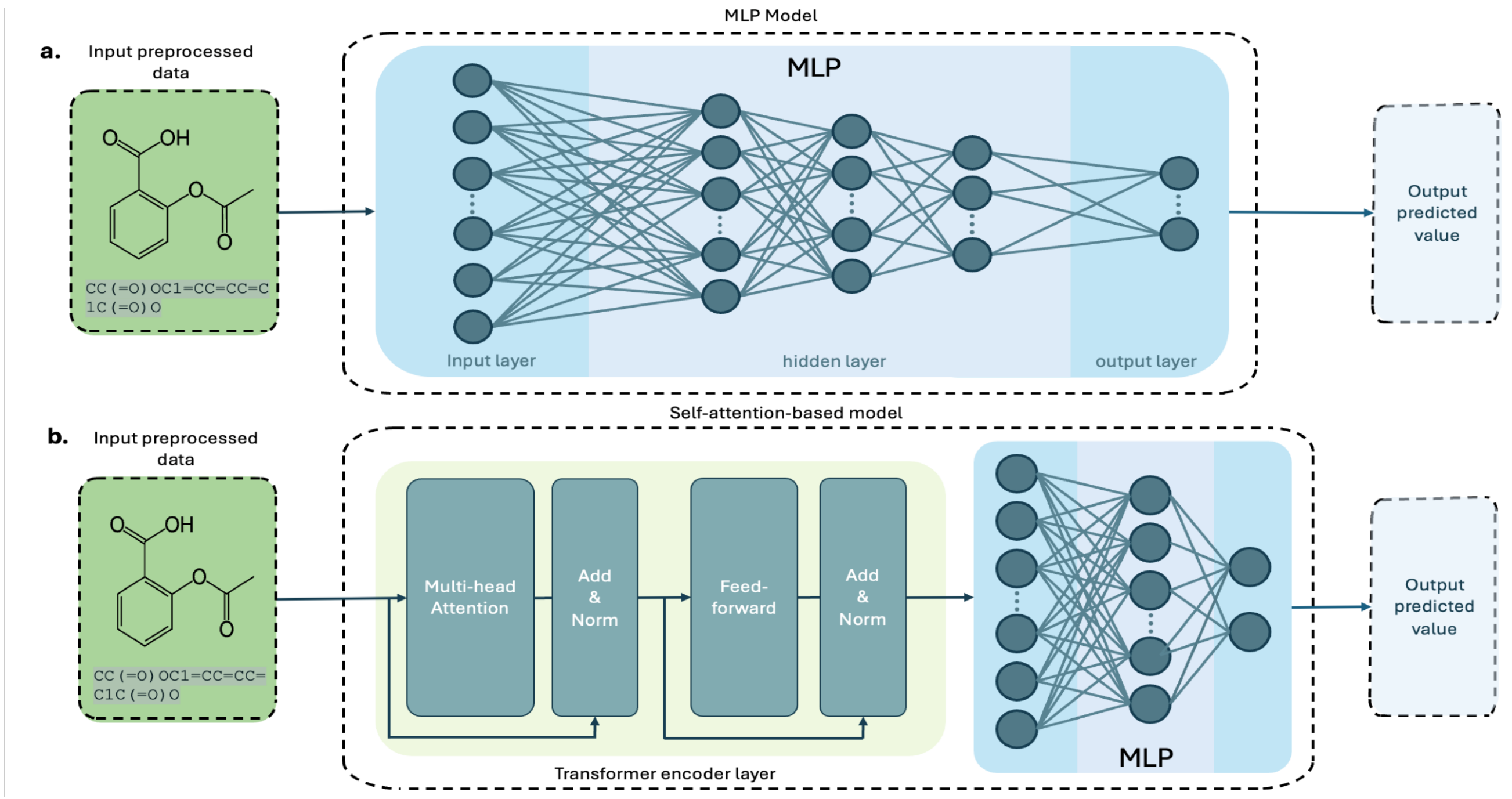

**Figure 1. Model architectures.** There are two different designs for the molecular property prediction model architectures. **(a)** Use the multi-layer perceptron (MLP) model to evaluate molecule embedding methods based on the model performance toward seven well-known datasets. **(b)** Applying the Transformer encoder layers combined with MLP to evaluate molecule embedding methods based on the model performance toward seven well-known datasets.

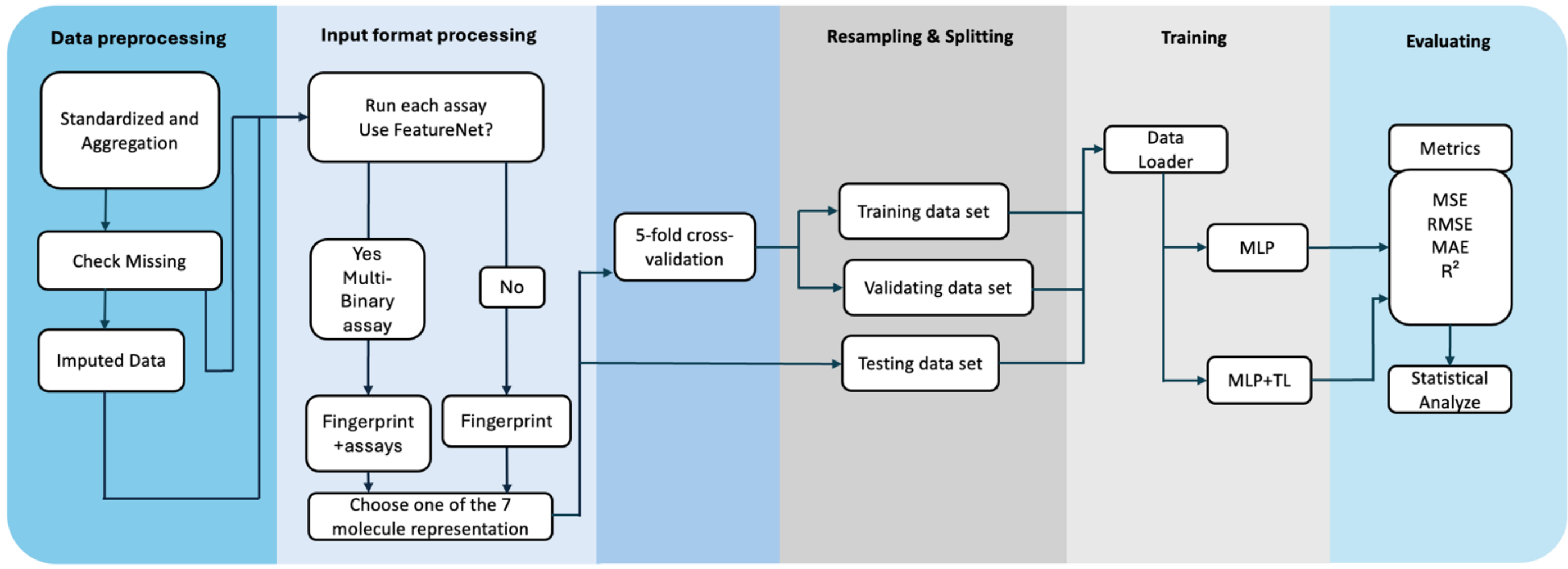

**Figure 2. Workflow of regression models and data processing design for the study.** The pipeline consists of five main steps: data cleaning and fingerprint standardization, optional transformation of the input format using FN, data resampling and splitting, model training, and calculation of performance metrics: Root Mean Squared Error (RMSE), Mean Squared Error (MSE), Mean Absolute Error (MAE), and Coefficient of Determination(R-squared)

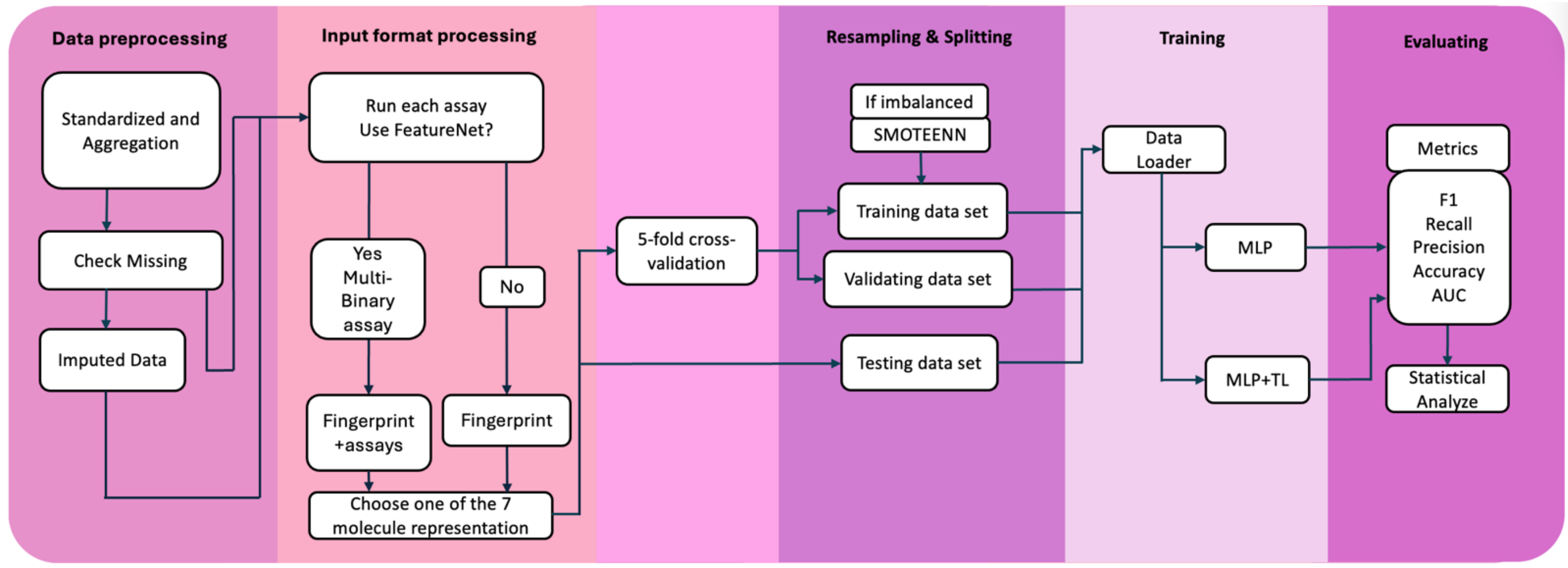

**Figure 3. Workflow of classification models and data processing design for the study.** The pipeline consists of five main steps: data cleaning and fingerprint standardization, optional transformation of the input format using FN, data resampling and splitting, model training, and calculation of performance metrics: F1, Recall, Precision, Accuracy, Area Under the Receiver Operating Characteristic Curve (AUC). If the training set is imbalanced after data splitting, an additional SMOTEENN step is applied.

# Materials and Methods

## Molecular encoding methods, standardization, and pre-processing

We implemented a comprehensive preprocessing pipeline for molecular bioactivity datasets by standardizing raw SMILES strings and aggregating duplicate compound annotations into a single consensus record. Specifically, we first removed invalid molecular strings, then applied a multi-step chemical standardization workflow, including hydrogen removal, metal disconnection, fragment cleanup, charge and radical neutralization, structural normalization, and tautomer canonicalization using RDKit and MolVS utilities; to further improve canonical consistency, we additionally performed an international chemical identifier (InChI) [34] round-trip conversion and preserved alternative outputs when discrepancies arise. After standardization, the pipeline filtered out non-organic or chemically ambiguous multi-fragment entries, retained only acceptable organic representations, and groups recorded by identical standardized SMILES. For each molecular group, labels were consolidated via majority voting (active vs. inactive), with ties marked as missing values to avoid introducing arbitrary bias. Finally, the processed dataset was exported with traceable identifiers linking each aggregated molecule back to its original row indices, thereby producing a cleaner, deduplicated, and analytically robust input for downstream modeling and evaluation. This workflow was adapted from a previously published molecular preprocessing strategy [35] and further extended in this study.

Molecular representations SMARTS and PubChem need to be precomputed before using in the pipeline. SMARTS were processed using natural language–inspired methods to generate machine-readable embeddings. SMILES were first converted into SMARTS patterns, which were treated as textual sequences describing substructure rules. These SMARTS strings were tokenized using a regular expression that separated chemical symbols based on structural delimiters such as -, \, /, =, #, ~, @, and :, thereby isolating atomic and bonding elements. After tokenization, a bag-of-words representation was constructed and reweighted with term frequency–inverse document frequency (TF-IDF) [36, 37], producing SMARTS embeddings that emphasized discriminative structural features. For each SMILES string, it queried PubChem through pubchempy, attempted to retrieve the corresponding compound record, and extracted the CACTVS/PubChem fingerprint as a binary feature vector; if the fingerprint was missing or any retrieval/parsing exception occurs, it substituted an all-zero vector of length 881 to preserve array shape consistency. We saved a compressed .npz file containing the fingerprint matrix, aligned SMILES list, and metadata (version, timestamp, dataset identifier, fingerprint type, sample count, and error statistics), thereby providing a traceable file which was suitable for using in our research pipeline.

## Molecular property datasets used for molecule encoding methods investigation

We leveraged seven benchmark molecular datasets that span a variety of molecular activities, including binding affinity, bioassay response, toxicity, and adverse reactions (Table 1). Those datasets would be used to train two models shown in Figures 1a-b, respectively. The BACE-1 [38] dataset focuses on the outcomes of molecule-human protein binding, particularly targeting the beta-secretase 1 enzyme. This dataset uses IC50 values to classify compounds as active or inactive, with inactive compounds defined by an IC50 ≤ 100 nM. The BBBP [39] dataset quantifies the ability of chemicals to penetrate the BBB, classifying compounds as either BBB+ or BBB-. The ClinTox [40] dataset integrates drug properties to identify clinical toxicity risks effectively, distinguishing between toxic and non-toxic compounds based on clinical trial outcomes. ToxCast [67] is a high-throughput in vitro toxicology dataset designed to characterize the biological activity of thousands of chemicals across a wide range of molecular targets and cellular pathways. The MUTAG [41] dataset consists of 188 aromatic and heteroaromatic nitro compounds labeled by their mutagenic effects in *Salmonella typhimurium*. This classification dataset is used to predict mutagenicity, helping identify compounds that may pose genetic risks. The SIDER [42] is a multi-label classification dataset, which is essential for predicting adverse drug reactions (ADRs), as each drug can belong to multiple ADR categories. The ESOL (Estimating Solubility) [43] dataset is a widely used benchmark dataset in cheminformatics for predicting the aqueous solubility of organic compounds. It consists of 1,128 molecules, each labeled with its experimentally determined water solubility (log solubility values in mol/L).

The external dataset used for BBBP prediction was B3DB [44], a chemically diverse molecular database compiled from 50 published data sources for BBB permeability research. It was developed to address the limitation that BBB permeability studies are often constrained by the small size and limited chemical diversity of existing training datasets. For ESOL prediction, the external datasets used were the AQUA and PHYS (A+P) datasets curated by Meng et al. [45]. Specifically, AQUA comprises experimental aqueous solubility data compiled from earlier public sources reported by Huuskonen[46] and Tetko [47], and PHYS is a curated subset of the PHYSPROP database processed through an automated KNIME workflow [45, 48]. These datasets provide a comprehensive foundation for molecular property predictions, supporting various aspects of drug discovery, safety assessment, and pharmacological research through classification and regression models.

**Table 1. Introduction of nine datasets including training sets and external validation sets.**

| **Dataset** | **BACE [38]** | **BBBP [39]** | **ClinTox [40]** | **ToxCast [67]** | **MUTAG [41]** | **SIDER [42]** | **ESOL** [43] | **B3DB [44]** | **A+P [44]** |
|---|---|---|---|---|---|---|---|---|---|
| **Size** | 1513 | 2050 | 1483 | 8597 | 188 | 1427 | 1128 | 7807 | 3321 |
| **Y-label** | active vs. inactive | penetration of the BBB | toxic and non-toxic | Various biological assay results | mutagenic vs. non-mutagenic | Side effects | log solubility values in mol/L | penetration of the BBB | log solubility values in mol/L |
| **Problem type** | Binary classification | Binary classification | Binary classification | Multi-label classification | Binary classification | Multi-label classification | Regression | Binary classification | Binary classification |

A+P: AQUA and PHYS dataset

## ToxCast curation

We used datasets from MoleculeNet [49] and MoCL [50] as reference benchmarks. However, most of these datasets were established several years ago and are no longer actively maintained. For instance, the latest publicly available release of SIDER is version 4.1 [42], which was released on October 21, 2015. By contrast, ToxCast remains actively expanded and provides integrated assay information from multiple sources, including Tox21. Accordingly, we decided to curate an updated ToxCast-based dataset ourselves from the latest available resources. We implemented an end-to-end, API-driven curation pipeline to construct a ToxCast binary activity matrix using the U.S. EPA CompTox services [67]. First, the complete assay catalog (AEIDs) was retrieved, and assay-level bioactivity records were systematically collected via the by-aeid endpoints. The assay outputs were subsequently standardized to reduce numerical instability and boundary-related artifacts. Specifically, small negative values close to zero (e.g., -1e-12), which were typically attributable to floating-point error rather than true negative activity, were reset to 0.0. Similarly, values slightly above 1 (e.g., 1.0000000001) were truncated to 1.0. Values that were clearly negative (i.e., less than -1e-6) were treated as missing. After cleaning, activity labels were binarized using a configurable threshold of $\geq 0.90$ [51]. For each AEID, assay-level records were compiled with corresponding chemical identifiers and activity annotations. The pipeline then queried the CompTox Chemical API to map each DTXSID to its corresponding SMILES representation, and records lacking valid chemical structures were excluded. When duplicate records were observed for the same molecule within the same assay, they were merged using a positive-dominant rule, such that the merged record was labeled active if any duplicate entry had a binary activity value of 1. The curated data were then consolidated into an assay-by-molecule matrix for downstream model training.

**Molecular property dataset pre-processing**

To ensure a meaningful and consistent evaluation across classification and regression tasks, we carefully selected target labels and applied dataset-specific preprocessing strategies tailored to the characteristics of each benchmark dataset. Since our comparison was based on single-task setting, we needed to select one label from multi-label datasets for model training. In the SIDER dataset, “Nervous system disorders” was selected as the target label because it owned the most balanced class. In parallel, we selected NVS_ADME_hCYP2C19 from ToxCast because a previous study has suggested that the labels retained after filtering generally showed better predictive performance than other ToxCast labels [8]. For the ToxCast dataset, assays were filtered to remove those with insufficient data labels for robust model training and evaluation. After chemical standardization and aggregation, only assays with at least 50 toxic and 50 non-toxic labels were retained [35]. To address the class imbalanced issue in datasets where the minority class proportion was below 0.3, we applied the SMOTEENN [52, 53] technique for oversampling. For the regression data, ESOL dataset, we applied `StandardScaler` [53] to the target variable (measured log solubility in mols per litre), fitting the scaler exclusively on the training set to prevent data leakage. The resulting scaler was then used to transform the training and validation labels before model training. After inference, both the predicted values and the ground-truth labels were restored to the original scale via `inverse_transform`, upon which the final performance metrics were computed. These preprocessing and label-selection procedures were designed to improve data quality, reduce bias caused by imbalance or scale differences, and provided a fair basis for downstream model comparison.

**Systematic investigation on molecular encoding methods for property predictions from neural network to the Transformer-encoder-based model**

**Multi-layer perceptron (MLP)**

The basic building block of an MLP [9] is the neuron, which is mathematically defined as:

$$z_j = \sum_{i=1}^{n} w_{ij} x_i + b_j \qquad (1)$$

where $z_j$ is the weighted sum of inputs for neuron $j$. $w_{ij}$ represents the weight from input $i$ to neuron $j$. $x_i$ is the input value. $b_j$ is the bias term for neuron $j$. $n$ indicates the total number of neurons for the certain neural network layer. The output of each neuron is shown in (2), which is obtained by applying an activation function [26].

$$\phi : \ a_j = \phi(z_j) \qquad (2)$$

Common activation functions include LeakyReLU [54] shown as below:

$$LeakyReLU(z) = \begin{cases} \alpha z, z \le 0 \\ z, z > 0 \end{cases} \qquad (3)$$

For the regression task, we will choose the root mean squared error (RMSE) to be a loss function as below:

$$L(y_i, \hat{y}_i) = \sqrt{\frac{1}{N}\sum_{i=1}^{N}(y_i - \hat{y}_i)^2} \qquad (4)$$

where $N$ is the number of samples, and $y_i$ is the $i$-th actual value. $\hat{y}_i$ indicates the $i$-th predicted value. Here, we want to obtain the optimal network parameter set $\theta^*$ to achieve the minimization of the loss function as follows:

$$\theta^* = arg \min_{\theta} L(\theta) \qquad (5)$$

which could be achieved by the backpropagation algorithm [9]. The forward and backward passes help us compute the gradient efficiently. In the MLP, the update of network parameter set, including weight ($w_j$) and bias ($b_j$) for the $k$-th times, is depicted in the following form:

$$\theta^k = \theta^{k-1} - \eta \nabla L(\theta^{k-1}) \qquad (6)$$

where $\eta$ is the learning rate. $\theta^{k-1}$ and gradient are given as below:

$$\theta^{k-1} = \begin{bmatrix} w_1 \\ \cdots \\ w_j \\ b_1 \\ \cdots \\ b_j \end{bmatrix}, \quad \nabla L(\theta^{k-1}) = \begin{bmatrix} \frac{\partial L(\theta^{k-1})}{\partial w_1} \\ \cdots \\ \frac{\partial L(\theta^{k-1})}{\partial w_j} \\ \frac{\partial L(\theta^{k-1})}{\partial b_1} \\ \cdots \\ \frac{\partial L(\theta^{k-1})}{\partial b_j} \end{bmatrix} \qquad (7)$$

**Transformer encoder-based model**

In the present model, the attention mechanism follows the same mathematical foundation as the Transformer [10], in which the core operation is defined as scaled dot-product attention:

$$\text{Attention}(Q,K,V) = \text{softmax}\left(\frac{QK^{\top}}{\sqrt{d_k}}\right)V \qquad (8)$$

Here, $Q$, $K$, and $V$ denote the query, key, and value matrices, respectively, and $d_k$ is the dimensionality of the key vectors. In the original Transformer, multi-head attention further extends this operation by projecting the same input into multiple subspaces and performing attention independently in each head:

$$MultiHead(Q,K,V) = Concat(head_1, \ldots, head_h)W^{O} \qquad (9)$$

$$\text{head}_i = \text{Attention}(QW_i^{Q}, KW_i^{K}, VW_i^{V}) \qquad (10)$$

Accordingly, the attention weights returned by the model correspond to the softmax-normalized matrix

$$A = \text{softmax}\left(\frac{QK^{\top}}{\sqrt{d_k}}\right) \qquad (11)$$

whose element $A_{ij}$ in A matrix quantifies how strongly the $i$-th query token attends to the $j$-th key token. Compared to the original Transformer, our model setting does not begin with a learnable token embedding derived from words or subwords. Instead, the input is a binary molecular fingerprint vector $x \in \{0,1\}^{L}$, where $L$ is the fingerprint length. For a batch of size $B$, the input tensor is $X \in \{0,1\}^{B\times L}$. The code first constructs an identity matrix $I_L \in \mathbb{R}^{L\times L}$, and each fingerprint bit $x_{b,i}$ is used to gate the corresponding one-hot basis vector. Thus, the initial representation is

$$E^{(0)}_{b,i,:} = x_{b,i}\, I_L[i,:] \qquad (12)$$

Equivalently, the full embedded tensor is:

$$E^{(0)} = \text{X}_{(:,:,\text{newaxis})} \odot I_L \qquad (13)$$

which has shape $(B, L, L)$. This means that each fingerprint position is treated as one token, and the feature dimension of each token is also $L$. If a bit is active, its token is initialized as the corresponding one-hot vector; if inactive, it becomes the zero vector. Therefore, unlike the original Transformer where token embeddings are dense and learned, our model setting starts from an explicit identity-based encoding of fingerprint positions, preserving direct interpretability of each bit from the very first layer. Furthermore, each encoder block consists of a multi-head self-attention (MHA) module followed by a position-wise feed-forward network, with residual connections and layer normalization applied after each sublayer. For an input tensor $E^{(\ell-1)} \in \mathbb{R}^{B\times L\times d_{\text{model}}}$ at the $\ell$-th layer, where $d_{\text{model}} = L$ in the present implementation, the self-attention mechanism first projects the input into query, key, and value spaces:

$$Q^{(\ell)} = E^{(\ell-1)} W_Q^{(\ell)},\ K^{(\ell)} = E^{(\ell-1)} W_K^{(\ell)}, V^{(\ell)} = E^{(\ell-1)} W_V^{(\ell)} \quad (14)$$

where $W_Q^{(\ell)}$, $W_K^{(\ell)}$, and $W_V^{(\ell)}$ are learnable projection matrices. For each attention head, the attention score between the $i$-th query token and the $j$-th key token is computed as

$$s_{ij}^{(\ell,\mathrm{h})} = \frac{\mathrm{q}_i^{(\ell,\mathrm{h})} \cdot \mathrm{k}_j^{(\ell,\mathrm{h})}}{\sqrt{d_k}} \quad (15)$$

where $d_k$ is the dimensionality of the key/query subspace. The attention weights are then obtained through a softmax operation over all key positions:

$$A_{ij}^{(\ell,\mathrm{h})} = \frac{\exp\,(s_{ij}^{(\ell,\mathrm{h})})}{\sum_{j'=1}^{L} \exp\,(s_{ij'}^{(\ell,\mathrm{h})})} \quad (16)$$

Thus, $\alpha_{ij}^{(\ell)}$ quantifies how strongly the $i$-th fingerprint bit attends to the $j$-th bit. Since the softmax is applied across the key dimension, the weights satisfy

$$\sum_{j=1}^{L} A_{ij}^{(\ell,\mathrm{h})} = 1 \quad (17)$$

The attention output for each token is then computed as the weighted sum of all value vectors:

$$\mathrm{z}_i^{(\ell)} = \sum_{j=1}^{L} \alpha_{ij}^{(\ell)}\, \mathrm{v}_j^{(\ell)} \quad (18)$$

In the multi-head setting, this process is conducted independently across $H$ heads, allowing the model to capture multiple patterns of inter-bit dependency. The returned attention tensor therefore has shape $(B, H, L, L)$, where each element represents the attention weight from one fingerprint position to another under a specific head. These weights can be directly used for interpretability analysis, such as constructing bit-to-bit attention heatmaps. In particular, by averaging across all heads in the last encoder layer, the model produces a head-averaged attention matrix $\bar{A} \in \mathbb{R}^{B\times L\times L}$, which is suitable for explanation on specific key that model capture:

$$\bar{A} = \frac{1}{H} \sum_{h=1}^{H} A_h^{(L_{enc})} \quad (19)$$

After self-attention, the block output is further refined by residual connection and layer normalization:

$$\tilde{E}^{(\ell)} = \mathrm{LayerNorm}\left(E^{(\ell-1)} + \mathrm{Dropout}(Z^{(\ell)})\right) \quad (20)$$

followed by a feed-forward network,

$$F^{(\ell)} = W_2^{(\ell)}\left(\mathrm{Dropout}\left(\sigma\left(W_1^{(\ell)}\tilde{E}^{(\ell)}\right)\right)\right) \quad (21)$$

where $\sigma(\cdot)$ denotes the activation function (ReLU [55]). A second residual connection and layer normalization are then applied:

$$E^{(\ell)} = \mathrm{LayerNorm}\left(\tilde{E}^{(\ell)} + \mathrm{Dropout}(F^{(\ell)})\right) \quad (22)$$

By stacking multiple such encoder blocks, the model progressively updates each fingerprint-bit representation based on both its own identity and its interactions with other bits, thereby learning higher-order dependencies that are not explicitly encoded in the original binary fingerprint. After the final encoder layer, the model converts the sequence of token representations into a fixed-length molecular representation using average pooling over the sequence dimension. If the final encoder output is $E^{(N)} \in \mathbb{R}^{B\times L\times L}$, then the pooled vector $p_b \in \mathbb{R}^L$ for the $b$-th sample is computed as

$$p_b = \frac{1}{L}\sum_{i=1}^{L} E_{b,i,:}^{(N)} \quad (23)$$

Accordingly, the mean pooled tensor has shape $P \in \mathbb{R}^{B\times L}$. This tensor $P$ serves as the direct input to the downstream MLP. Importantly, the MLP does not receive the original fingerprint vector $x$ itself; rather, it operates on a transformed global representation obtained after identity-based expansion, multi-layer self-attention, and sequence-level aggregation. Therefore, the final prediction can be summarized as

$$\hat{y}_b = MLP(p_b) = \mathrm{MLP}\left(\frac{1}{L}\sum_{i=1}^{L} \mathrm{E}_{b,i,:}^{(N)}\right) \quad (24)$$

This formulation indicates that the MLP input is a contextualized molecular representation in which each fingerprint bit has already been updated according to its learned interactions with all other bits.

**Multi-assay data imputation by Feature Net**

Motivated by the findings of Walter et al. [35], which demonstrated that imputation-based approaches such as Feature Net (FN) [35, 56, 57] substantially improve predictive performance on sparse multi-assay datasets (ToxCast), we extended this methodology to our study. The FN [35, 56, 57] framework operated in two stages. Firstly, a single-task model (XGBoost [58]) was trained for each assay using chemical features to predict missing labels; secondly, each assay was re-modeled (MLP/MLP+TL) by integrating chemical features with experimentally determined or imputed auxiliary assay labels as additional inputs. This design enabled the model to explicitly leverage related assays for improved prediction of the target endpoint. Based on this rationale, we applied the FN [35, 56, 57] framework to the multi-assay datasets SIDER, ClinTox, and ToxCast. For SIDER, all 26 binary adverse event labels were incorporated as auxiliary assay inputs. For ClinTox, the binary endpoint FDA_APPROVED was included as an auxiliary label for predicting the target toxicity-related outcome. For ToxCast, which comprised assays spanning diverse biological processes and targets, we incorporated 553 assays as auxiliary inputs. All auxiliary endpoints used in this study were binary (0/1) labels. The incorporation of FN [35, 56, 57] led to improve predictive performance across these multi-assay datasets (Tables S18, S21, and S22). Additionally, FN [35, 56, 57] is particularly advantageous for compounds with chemically dissimilar data, for cases with multiple auxiliary assay labels, and when experimental auxiliary labels are available as suggested in the previous study [35]. These considerations guided our integration of FN [35, 56, 57] into our modeling frameworks, thereby broadening the applicability domain and enhancing prediction reliability under data-sparse conditions.

**Training and Testing**

For model development, the dataset was first divided into a training-validation set and an internal test set at a ratio of 8:2. Within the training-validation set, five-fold cross-validation was performed to determine the optimal hyperparameters. After hyperparameter selection, the data were again split into training and validation subsets at a ratio of 8:2 for final model training. Early stopping was applied during training to prevent overfitting and improve generalization by terminating training when the validation performance no longer improved, thereby also reducing computational cost. For subsequent molecular-level prediction and interpretation, we used the optimal hyperparameters obtained from five-fold cross-validation. The target molecule to be interpreted was treated as the single test sample, while the remaining molecules were split into training and validation sets at a ratio of 8:2 for model training with early stopping. To evaluate model performance, different metrics were adopted for classification and regression tasks. For classification tasks, the area under the receiver operating characteristic curve (AUC-ROC) was adopted as the primary metric in five-fold cross-validation. Unlike the F1 score, AUC-ROC is threshold-independent and therefore provided a more comprehensive assessment of model performance across different decision boundaries. Owing to this property, it has been widely applied in biological and biomedical data analysis. In contrast, for the solubility prediction task, which was formulated as a regression problem on a continuous numerical dataset, RMSE was used as the principal evaluation metric because it is commonly employed in the literature and enables direct comparison with previous studies [45]. For BBBP external validation, since B3DB [44] contained molecules originating from BBBP, we first removed all molecules in B3DB that shared the same CID or InChI [34] with molecules in BBBP. We further observed that some molecules in B3DB still had the same compound names as those in BBBP. To avoid potential data leakage, these molecules were removed from B3DB as well. After this filtering process, the final size of B3DB was 4,368 molecules. In addition, after standardizing the molecules in the A+P datasets, which served as the external dataset for ESOL, we checked whether any molecules had the same SMILES strings as those in ESOL. If duplicated SMILES were identified, the corresponding molecules were removed from A+P. The remaining molecules from A+P were then combined to form the external dataset. After that, the combined size of A+P owned 1362 molecules. To assess whether the external dataset had a distribution comparable to that of the training dataset, visualization was performed (Figure S8a-b).

**Table 2. Summary of dataset preprocessing.** After filtering, the number of assays in ToxCast decreased from 1,501 to 555.

| Dataset | ToxCast | ClinTox | MUTAG | BBBP | SIDER | ESOL | BACE |
|---|---|---|---|---|---|---|---|
| **y number** | 1501→555 | 2 | 1 | 1 | 27 | 1 | 1 |
| **X number** | 9002 | 1483 | 188 | 2050 | 1427 | 1128 | 1513 |
| **after process X number** | 7787 | 1380 | 186 | 1943 | 1293 | 1117 | 1500 |
| **Number of X requiring imputation** | 7422 | 45 | none | 11 | 1 | 11 | 1 |
| **Input** | fingerprint+assays | fingerprint+assay | fingerprint | fingerprint | fingerprint+assays | fingerprint | fingerprint |
| **Representation assay** | NVS_ADME_hCYP2C19 | CT_TOX | label | p_np | 'Nervous system disorders' | ‘measured log solubility in mols per litre’ | Class |
| **Extra preprocessing** | filter 50/50 | | | | | | |
| **Use smoteenn** | None | Yes | None | Yes | Yes | numerical data | None |

y: the number of assays; and X: the number of molecules.

### Identifying representative substructures via attention score analyses

For MLP+TL model setting, each binary molecular fingerprint was first converted into a structured token matrix by masking an identity matrix with the input bits, so that active bits preserved their one-hot basis vectors while inactive bits became zero rows as shown in Equation (12) and (13). This matrix was then processed by stacked Transformer encoder layers. By averaging across all heads in the last encoder layer, the model produced a head-averaged attention matrix, which was used to explain the important structure the model capture. For each molecule, the attention weights were summed across all queries for each key, yielding a column-wise attention sum. The top 20 keys with the highest summed attention values were selected for each sample (Table S33), and the most frequently occurring bits across these rankings were subsequently summarized at the dataset level. The top 30 most frequent bits at the dataset level were further grouped into chemically interpretable categories representing major determinants of the target properties (Figure S9a-d). Additionally, when discussing important substructures related to BBBP, we used Morphine and Heroin as examples. For these two compounds, we calculated the mean attention weight for each of the 167 MACCS keys (the column of matrix) to further examine whether substructures absent from the fingerprint (encoded as

0) were systematically assigned lower attention weights by the MLP+TL model. The function of the mean attention weight on columns is shown as below:

$$col_{mean_k} = \left(\frac{1}{167}\right)\sum_{q=0}^{166} Attention[q,k] \qquad (25)$$

# Results

We focus on two neural network-based architectures (Figures 1a-b): MLP and MLP+TL trained on seven well-known molecular datasets to compare the performance of different molecular encoding methods. We further explore the model's internal attention weights as an interpretability-related signal for identifying potentially important features. Thus, this study aimed not only to assess predictive performance across molecular representations, but also to investigate whether an intermediate modeling strategy could provide a practical balance between predictive capability and interpretability. Overall, the MLP model achieved the best performance on most datasets (Figures 4a-g, Table 4). However, in most cases, its advantage over MLP+TL was not substantial, suggesting that MLP+TL remains a valuable alternative (Table 3, Figures S1-S7). Although MLP+TL did not consistently outperform MLP, it offers additional interpretability for analyzing property-related substructures and identifying important structural patterns associated with different molecular properties. Across datasets, the optimal fingerprint varied depending on the prediction task and model architecture, highlighting that fingerprint choice is a critical factor in molecular property prediction (Table 3).

**Comparative study of molecular property predictions based on multi-layer perceptron and Transformer encoder-based model**

Based on five-fold cross validation results, for the BACE dataset (Figure 4a, Table 4), FCFP achieved the highest average of AUC value in MLP (average of AUC=0.882 ± 0.017), whereas ECFP performed best in MLP+TL (average of AUC = 0.879 ± 0.025). On the BBBP dataset (Figure 4b, Table 4), FCFP performed best, with average AUC values of 0.891 ± 0.021 and 0.871 ± 0.015 for MLP and MLP+TL, respectively. For ClinTox (Figure 4c, Table 4), MACCS achieved the highest average AUC in the MLP model (average of AUC=0.925 ± 0.023). In the MLP+TL model, FCFP yielded the highest average value of AUC (average of AUC =0.917 ± 0.023), differing by approximately 0.01 from SMARTS; however, its performance on other evaluation metrics was weaker (Table S7). Therefore, considering overall performance rather than AUC alone, we regarded SMARTS as the most suitable fingerprint for the MLP+TL model on ClinTox. On the MUTAG dataset (Figure 4e, Table 4), MACCS achieved an average AUC value of 0.913 ± 0.026 with MLP, whereas ECFP reached average AUC value of 0.920 ± 0.046 with MLP+TL. On the SIDER dataset (Figure 4f, Table 4), SMARTS achieved strong and consistent performance in MLP and MLP+TL models with average AUC values of 0.916 ± 0.021 and 0.918 ± 0.013, respectively, indicating that it is the most appropriate fingerprint for this task. On the ToxCast dataset (Figure 4g, Table 4), RDKit with MLP achieved the best result owning average AUC value of 0.974 ± 0.004, whereas FCFP with MLP+TL reached average AUC value of 0.950 ± 0.005. For the ESOL dataset (Figure 4d, Table 4), the best average RMSE values were 0.986 ± 0.063 and 1.132 ± 0.034 obtained by RDKit with MLP and MLP+TL models. According to the previous study [45], the RMSE for solubility prediction is commonly reported in the range of 0.7-1.0; therefore, our results are within a reasonable range. Among all datasets, ESOL and ToxCast showed the largest differences between MLP and MLP+TL across evaluation metrics showing significant differences observed in 100% and 77% of the comparisons, respectively (Table 3, Figures S4 and S7). Notably, SMARTS and RDKit showed the largest performance variation across datasets, frequently ranking either first or last among the evaluated fingerprints. In contrast, FCFP, ECFP, and MACCS tend to provide more robust and consistent representations across most datasets and models, as they were more frequently ranked among the top three fingerprints, especially in the MLP model (Figures 4a-g). This pattern reinforces the notion that fingerprint selection is not merely a preprocessing choice, but a major determinant of downstream predictive performance.

**Table 3. Summary of experiment results on seven datasets.** ESOL and ToxCast demonstrated the most significant differences between MLP and MLP+TL across various evaluation metrics, with notable differences observed in 100% and 77% of the comparisons, respectively. SMARTS and RDKit exhibited the most substantial performance variation across datasets. In contrast, FCFP, ECFP, and MACCS tend to provide more reliable and consistent representations across most datasets and models.

| **Dataset** | **ToxCast** | **ClinTox** | **MUTAG** | **BBBP** | **SIDER** | **ESOL** | **BACE** |
|---|---|---|---|---|---|---|---|
| **Significant difference** | 77% | 20% | 20% | 0% | 3% | 92% | 20% |
| **MLP best fp** | RDKit | MACCS | MACCS | FCFP | SMARTS | RDKit | RDKit |
| **MLP+TL best fp** | FCFP | SMARTS | ECFP | FCFP | SMARTS | SMARTS | RDKit |

fp: fingerprint

**Table 4. Performance comparison of MLP and MLP+TL across benchmark and external validation datasets.** Average five-fold cross-validation result of AUC scores for six classification datasets (BACE, BBBP, ClinTox, MUTAG, SIDER, and ToxCast), average five-fold cross-validation result of RMSE for the ESOL regression dataset, and performance on two external validation datasets (B3DB, A+P).

| | ECFP | | FCFP | | MACCS | | PubChem | | RDKit | | SMARTS | |
|---|---|---|---|---|---|---|---|---|---|---|---|---|
| | MLP | MLP+TL | MLP | MLP+TL | MLP | MLP+TL | MLP | MLP+TL | MLP | MLP+TL | MLP | MLP+TL |
| BACE | 0.879 ± 0.029 | **0.879 ± 0.025** | **0.882 ± 0.017** | 0.859 ± 0.025 | 0.868 ± 0.033 | 0.842 ± 0.032 | 0.849 ± 0.025 | 0.570 ± 0.106 | 0.881 ± 0.022 | 0.870 ± 0.019 | 0.762 ± 0.013 | 0.722 ± 0.034 |
| BBBP | 0.845 ± 0.029 | 0.862 ± 0.041 | **0.891 ± 0.021** | **0.871 ± 0.015** | 0.855 ± 0.013 | 0.851 ± 0.012 | 0.746 ± 0.050 | 0.751 ± 0.042 | 0.880 ± 0.027 | 0.814 ± 0.060 | 0.849 ± 0.028 | 0.816 ± 0.035 |
| ClinTox | 0.836 ± 0.036 | 0.801 ± 0.033 | 0.921 ± 0.051 | **0.917 ± 0.023** | **0.925 ± 0.023** | 0.903 ± 0.019 | 0.792 ± 0.082 | 0.805 ± 0.089 | 0.814 ± 0.032 | 0.674 ± 0.076 | 0.909 ± 0.045 | **0.916 ± 0.035** |
| ESOL | 1.178 ± 0.045 | 1.354 ± 0.082 | 1.105 ± 0.079 | 1.263 ± 0.094 | 1.008 ± 0.040 | 1.147 ± 0.044 | 1.041 ± 0.082 | 1.183 ± 0.099 | **0.986 ± 0.063** | **1.132 ± 0.034** | 1.274 ± 0.054 | 1.521 ± 0.103 |
| MUTAG | 0.906 ± | **0.920 ±** | 0.896 ± | 0.853 ± | **0.913 ±** | 0.806 ± | 0.897 ± | 0.876 ± | 0.878 ± | 0.879 ± | 0.830 ± | 0.752 ± |

|  | 0.043 | **0.046** | 0.069 | 0.059 | **0.026** | 0.086 | 0.047 | 0.041 | 0.057 | 0.040 | 0.073 | 0.062 |
|---|---|---|---|---|---|---|---|---|---|---|---|---|
| SIDER | 0.882 ± 0.014 | 0.908 ± 0.020 | 0.907 ± 0.021 | **0.918 ± 0.009** | 0.881 ± 0.032 | 0.880 ± 0.032 | 0.860 ± 0.035 | 0.881 ± 0.025 | 0.772 ± 0.053 | 0.744 ± 0.067 | **0.916 ± 0.021** | **0.918 ± 0.013** |
| ToxCast | 0.959 ± 0.004 | 0.929 ± 0.009 | 0.971 ± 0.001 | **0.950 ± 0.005** | 0.955 ± 0.002 | 0.939 ± 0.009 | 0.971 ± 0.001 | 0.949 ± 0.010 | **0.974 ± 0.004** | 0.883 ± 0.011 | 0.907 ± 0.010 | 0.814 ± 0.046 |
| B3DB | 0.8035 | 0.7993 | 0.8224 | 0.6619 | 0.8105 | **0.8111** | 0.7167 | 0.7058 | **0.8308** | 0.7474 | 0.7394 | 0.7488 |
| A+P | 1.3849 | 1.4111 | 1.456 | 1.5274 | 1.2557 | 1.4211 | **1.1232** | **1.3047** | 1.3338 | 1.4542 | 1.3474 | 1.5911 |

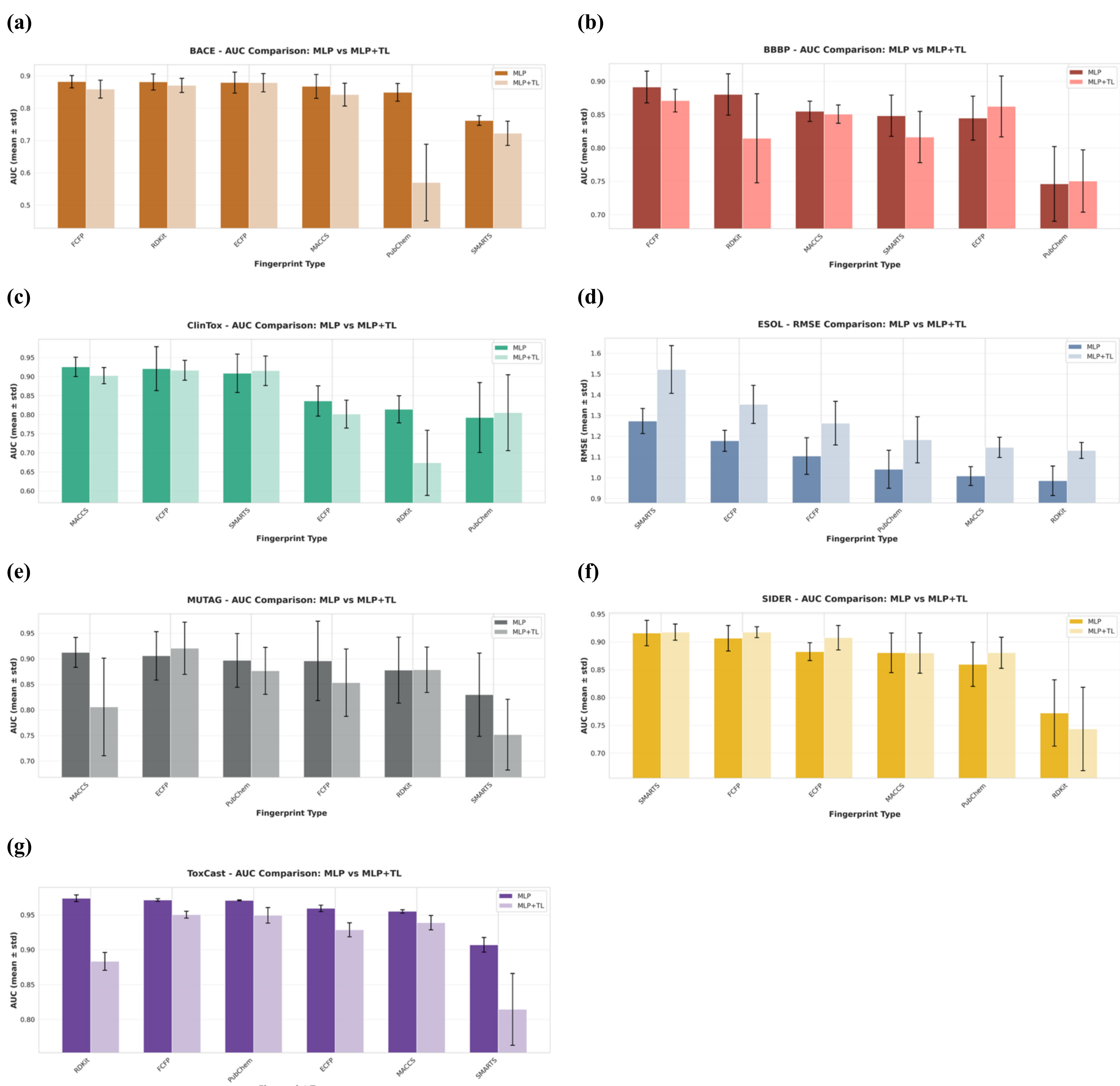

**Figure 4. Bar plots of five-fold cross-validaiton results of AUC scores showing the ranking of fingerprint methods based on the performance of the MLP and MLP+TL models across different datasets**: (a) BACE, (b) BBBP, (c) ClinTox, (d) ESOL, (e) MUTAG, (f) SIDER, (g) ToxCast.

**External datasets validation for BBB penetration and solubility predictions based on multi-layer perceptron and Transformer encoder-based model**

To the validation results on B3DB for BBB penetration prediction, RDKit, FCFP and MACCS consistently demonstrated superior performance in the MLP model (Table 4, Table S23). Conversely, the performance of SMARTS and PubChem experienced a significant decline, with both scores dropping below 0.8. In the MLP+TL model, the performance of FCFP further decreased to 0.66 (Table 4, Table S23). On the A+P external validation set for solubility prediction, the RDKit in the MLP+TL model exhibited the most substantial increase in average RMSE, reaching 1.45, resulting in a drop in ranking from first to fourth (Table 4, Table S24). Despite the presence of some outliers in the two datasets when using MLP+TL model, which suggested a weaker generalization ability. The ranking order for all molecular representations remained largely consistent across both datasets in MLP model, indicating an overall stable performance trend (Tables S23-24).

**Physicochemical feature groups identification for BBB penetration prediction based on attention score analyses**

Prior studies have shown that BBB permeability is largely governed by lipophilicity, polarity, hydrogen-bonding capacity, ionization state, and flexibility [59, 60]; therefore, the substructures identified in BBBP were grouped according to these properties. The top-ranked MACCS keys can be organized into four chemically meaningful groups, which together reflected the major determinants of BBB permeability (Table 5, Figures S9a-b). The first group comprised hydrophobic features, which were generally favorable for passive BBB permeation because they increased hydrophobic surface area, reduced flexibility, and helped molecule across the lipid membrane. Representative examples included MACCS key 163 (6M Ring). The second group consisted of oxygen-containing polar substructures, which were usually unfavorable for BBB penetration because they increased polarity, hydrophilicity, and hydrogen-bond donor/acceptor capacity. A representative example was MACCS key 139 (OH), highlighting detrimental effect of hydroxyl groups on passive membrane diffusion. The third group included nitrogen-containing and ionizable substructures**,** as strong positive charge, or high ionization reduced passive BBB permeability. An example was MACCS key 121 (N Heterocycle). The fourth group was composed of topology- and linker-related patterns, which did not correspond to a single functional group but reflected scaffold connectivity, ring topology, and molecular flexibility. An example was MACCS key 98 (QAAAAA@1). In summary, these four groups suggested that the model captured chemically meaningful features associated with BBB permeation, including lipophilicity, polarity, hydrogen-bonding capacity, and ionization state. Moreover, the PubChem top-ranked substructures were broadly consistent with the conclusions from the MACCS analysis. In particular, the PubChem patterns were enriched in hydrophobic fragments, aromatic and six-membered ring substructures, and oxygen-, nitrogen-containing features. Compared with the MACCS representation, which highlighted hydroxyl-related patterns more explicitly, the PubChem fingerprint provided a broader substructure and therefore captured the overall chemical space in a less functional-group-specific manner.

**Table 5. Classification of the top-ranked MACCS and PubChem keys for BBBP prediction into four chemically meaningful groups.** These groups were defined according to the major physicochemical factors known to influence blood–brain barrier permeability, including hydrophobicity, polarity, hydrogen-bonding

ability, ionization.

| Group | MACCS key structures | PubChem key structures |
|---|---|---|
| **Group 1. Hydrophobic, ring substructure** | - Aromatic<br>- 6M Ring<br>- 6M ring > 1<br>- ACH2AAACH2A<br>- ACH2AACH2A<br>- CH3 > 1<br>- XA(A)A | - C:C-C-C-C<br>- C-C-C-C-C-C-C<br>- C(:C)(:C)<br>- >= 2 any ring size 6<br>- C-C(C)-C-C<br>- C:C-C-C:C<br>- C-C-C(C)-C-C<br>- C(-H)(=C)<br>- C(~C)(~C)<br>- >= 1 aromatic ring<br>- C(~C)(~C)(~C)<br>- >= 1 any ring size 6 |
| **Group 2. Oxygen-containing polar substructures** | - O<br>- OH<br>- QHAAAQH<br>- C=O<br>- OACH2A<br>- OQ(O)O<br>- A$A!O > 1<br>- CSO | - C(-C)(=O)<br>- >= 1 O<br>- O-C:C-C-C<br>- O(~C)(~H) |
| **Group 3. Nitrogen-containing and ionizable substructure** | - N Heterocycle<br>- AN(A)A<br>- ACH2N<br>- CH2QCH2<br>- NCO<br>- NAAO<br>- NAO<br>- NN<br>- N=A<br>- CHARGE | - N-C-C-C:C<br>- C-N<br>- N-H<br>- N(~C)(~C)<br>- N(-C)(=C) |
| **Group 4. Topology- and linker-related patterns** | - Heterocycle<br>- QAAAAA@1<br>- A!A$A!A<br>- A!CH2!A<br>- 8M Ring or larger | - C-S<br>- N-O<br>- C-Cl<br>- C-Se<br>- C-As<br>- C-I<br>- C-Br<br>- O-C-C-N<br>- C-Si |

**Substructure grouping identification in mutagenicity prediction based on attention score analyses**

In contrast to BBBP, the MUTAG dataset is more naturally interpreted in terms of toxicophore families rather than physicochemical determinants of membrane permeation. Based on previous mutagenicity studies, the key structures in MUTAG are expected to reflect major mutagenicity-related features, including direct genotoxic alerts, nitrogen-containing substructuress, aromatic or conjugated scaffolds, and substituent or topological contexts that modulate alert strength [41, 61-63]. As a result, its top-ranked structures are expected to cluster into a smaller number of chemically coherent mutagenicity-related groups (Table 6, Figures S9c-d). Importantly, not every top-ranked fragment in MUTAG should be interpreted as a direct toxicophore. Some features correspond to direct alerts, whereas others represent scaffold or background contexts that modulate the strength or expression of those alerts. For example, a fragment such as Ring does not by itself constitute a mutagenicity alert, but it provides stabilization that support the activity of nitro- or aromatic amine-related toxicophores. Therefore, the structural interpretation of MUTAG should be understood at two levels: first, direct mutagenicity alerts, such as nitro, N–O, and aromatic amine substructures; and second, scaffold features, such as aromatic rings, fused or conjugated systems, and substituent topology, which influence electronic distribution and reactive intermediate formation. This distinction is chemically important because mutagenicity often depends not only on the presence of an alerting functional group, but also on the structural environment that determines its activation and reactivity.

At a finer level, the top-ranked MACCS and PubChem structures in MUTAG can be organized into five groups (Table 6, Figures S9c-d). The first group contained direct nitro/N–O-related mutagenicity alerts, such as MACCS key 63 N=O, which were consistent with the role of nitro reduction and N–O-containing intermediates in mutagenic activation. The second group contained arylamine or aniline-like nitrogen substructures, such as PubChem key 737 Cc1cc(N)ccc1, which were chemically relevant because aromatic amines can undergo metabolic activation through N-hydroxylation and subsequently generated DNA-reactive intermediates. The third group consisted of aromatic, heteroaromatic, and conjugated scaffold features, exemplified by MACCS key 165 Ring, which reflected the planar and resonance-stabilized backgrounds that supported mutagenic intermediate formation. The fourth group included oxygen-content and charge-related patterns, such as MACCS key 185 CHARGE, which reflected oxidation state or electronic environments associated with reactive functional groups. The fifth group contained alkyl, alkenyl, and substituent-topology substructures, exemplified by PubChem key 634 C-C-C:C-C, which were better interpreted as alert-modulating features than as direct mutagenicity alerts. Overall, these grouped patterns indicated that the model was not simply attending to random fragments, but rather to a set of chemically coherent substructures that aligned with known mechanisms underlying nitroaromatic and aromatic amine mutagenicity. In summary, BBBP and MUTAG differed not only in prediction task but also in the most appropriate mode of interpretation. For BBBP, grouped feature analysis mainly reflected the physicochemical requirements for crossing the BBB, such as lipophilicity, limited hydrogen bonding and controlled ionization. For MUTAG, grouped feature analysis instead reflected the presence of mutagenicity-associated toxicophores and the structural patterns that enabled their metabolic activation and DNA reactivity. In conclusion, our MLP+TL model can offer biological insight through substructure-level interpretation across different datasets.

**Table 6. Classification of the top-ranked MACCS and PubChem keys for MUTAG prediction into five chemically meaningful groups.** These groups were defined according to previously reported mutagenicity-

related structural themes, including direct toxicophore alerts, aromatic amine- and nitro-related substructures, conjugated aromatic scaffold, charge/oxygen-associated features, and substituent-topology effects.

| Group | MACCS key structures | PubChem key structures |
|---|---|---|
| **Group 1. Direct alerts (nitro / N-O)** | - N=O<br>- N-O<br>- ON(O)C<br>- NAAAO<br>- QNQ<br>- QN<br>- QO | - N(~O)(~O)<br>- O=N-C:C<br>- O-N-C-C |
| **Group 2. Arylamine or aniline-like nitrogen substructures** | - N<br>- N > 1<br>- NH<br>- AN(A)A<br>- NA(A)A<br>- NAAN<br>- AQ(A)A | - N-C:C:C-C<br>- C(~N)(:C)<br>- C(~N)(:C)(:C)<br>- Cc1cc(N)ccc1<br>- CC1CC(N)CCC1<br>- >= 2 N |
| **Group 3. Aromatic, heteroaromatic, and conjugated scaffold features** | - `Ring`<br>- `6M Ring`<br>- `6M ring > 1`<br>- `Aromatic`<br>- `Aromatic Ring > 1`<br>- 5 M ring<br>- A$A($A)$A<br>- A$A!N<br>- A!A$A!A<br>- Anot%A%Anot%A<br>- Nnot%A%A | - C:C-C:C<br>- >= 3 aromatic rings<br>- >= 2 aromatic rings<br>- >= 1 any ring size 6<br>- >= 1 any ring size 5<br>- C:C-C-C:C<br>- C:C-C-C-C |
| **Group 4. Oxygen-content and charge-related patterns** | - CHARGE<br>- QQ<br>- O<br>- O > 2<br>- O > 3 (&...) Spec Incomplete | >=2O |
| **Group 5. Alkyl, alkenyl, and substituent-topology substructures** | | - C-C-C:C-C<br>- [#1]-C=C-[#1]<br>- C(~C)(:C)(:C)<br>- C-C-C-C-C<br>- C-C-C-C-C-C<br>- C(-C)(-H)(=C)<br>- C-C-C(C)-C-C |
| | | - C-C(C)-C-C<br>- C-C-C=C<br>- C-C=C-C=C<br>- C=C-C-C-C<br>- C-C-C=C-C<br>- C(-C)(-C)(=C) |

**Comparing identified substructure between Morphine and Heroin based on attention score analyses**

The comparison between Morphine and Heroin highlighted the importance of hydroxyl groups in BBB permeability prediction. The two molecules share the same core scaffold but differ in functional group composition (Figure 6). This comparison was primarily associated with the second group, because the key difference lied in the presence or masking of hydroxyl-related polar substructures. Free hydroxyl groups, as present in Morphine, increase molecular polarity and hydrogen-bond donor capacity, both of which are generally unfavorable for BBBP. By contrast, Heroin, a diacetylated derivative of Morphine, exhibits increased lipophilicity and therefore crosses the BBB more efficiently. As a result, Heroin penetrates the BBB more readily, suggesting that hydroxyl-group-related features are key structural determinants of BBB permeability [64].

Analysis of the most important substructures revealed a clear difference between the two molecules. An OH-related substructure was identified among the top 20 most important features for Morphine, whereas no such feature appeared among the top 20 features for Heroin (Table S33). This difference was also reflected in the attention heatmap in Figures 5a and 5c. In the MACCS fingerprint, key 139 corresponds to OH. In morphine, free hydroxyl groups increase molecular polarity and hydrogen-bond donor capacity, both of which tend to hinder BBB permeation. Compared with other columns, Morphine showed a darker signal in column 139, indicating higher attention weights were assigned to this feature (Figures 5b, d). A quantitative comparison of the attention weights associated with key 139 further supported this observation. In Morphine, the maximum attention weight was 0.01199, originating from query 98, whereas in Heroin, the maximum value was 0.00711, originating from query 122 (Figure 7, Table S34). Morphine also exhibited a more heterogeneous attention pattern; however, heroin displayed a relatively uniform distribution (Figure 7).
Heroin do not have key 139 (OH), which is mark 0 in MACCS. Subsequently, we further would like to examine whether absent substructure encoded as 0 leading to lower attention weights assigned by MLP+TL systematically. To address this, we calculated the mean attention weight for each of the 167 keys in Morphine and Heroin (Figures 8a-b). Notably, the results showed that some keys labeled as 1 still had lower column mean attention weights than keys labeled as (Figure 8b). In addition, for Heroin, all keys labeled as 0 had column mean attention weights above the global mean of the column-wise attention weights (Figure 8a). In other words, the keys labeled as 0 were not systematically ignored by the model. Together, these results indicated that the model is able to capture the hydroxyl-related structural difference between Morphine and Heroin. Moreover, the analysis suggested that attention was not restricted only to keys encoded as 1, supporting a more distributed pattern of feature weighting.

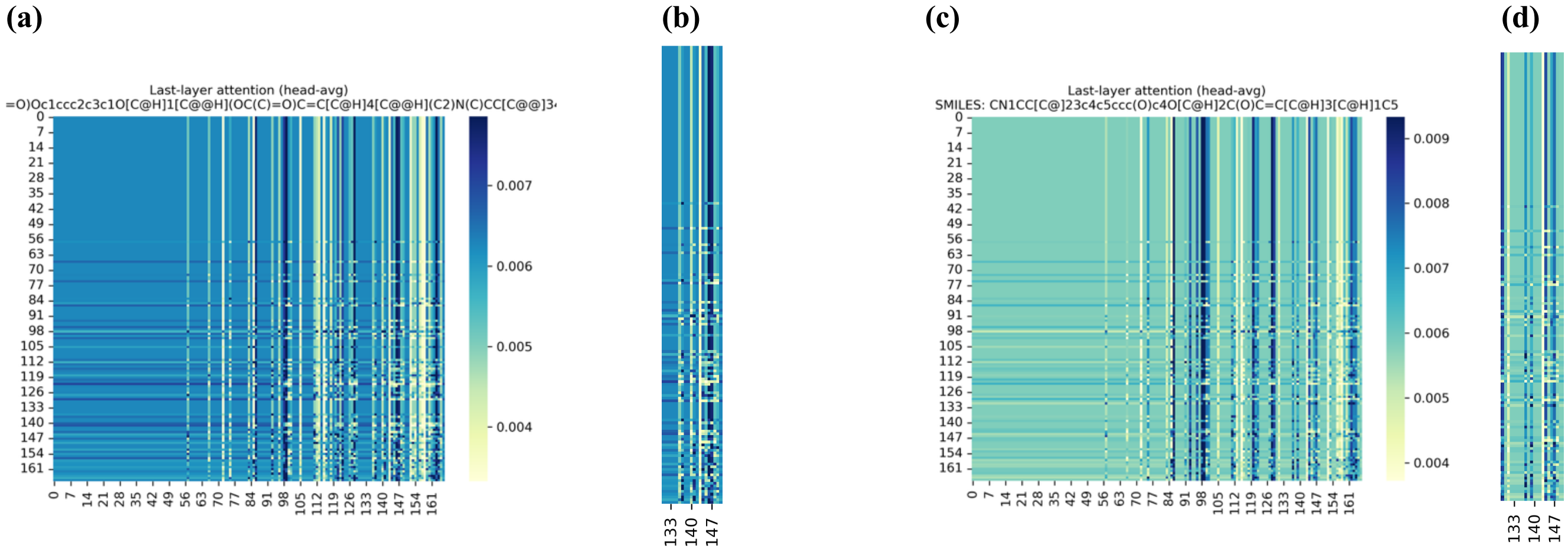

**Figure 5. (a) The attention score heatmap of Heroin.** Key 139 appears relatively light compared with other columns, indicating weaker attention to this feature in heroin. **(b) Zoomed-in view of key 139 in Heroin. (c) The attention score heatmap of Morphine.** Key 139 appears relatively dark compared with other columns, suggesting that the model assigns greater attention to the OH-related feature in morphine. **(d) Zoomed-in view of key 139 in Morphine.** The y-axis represents queries, and the x-axis stands for keys. In the MACCS fingerprint, key 139 corresponds to a hydroxyl (OH) group.

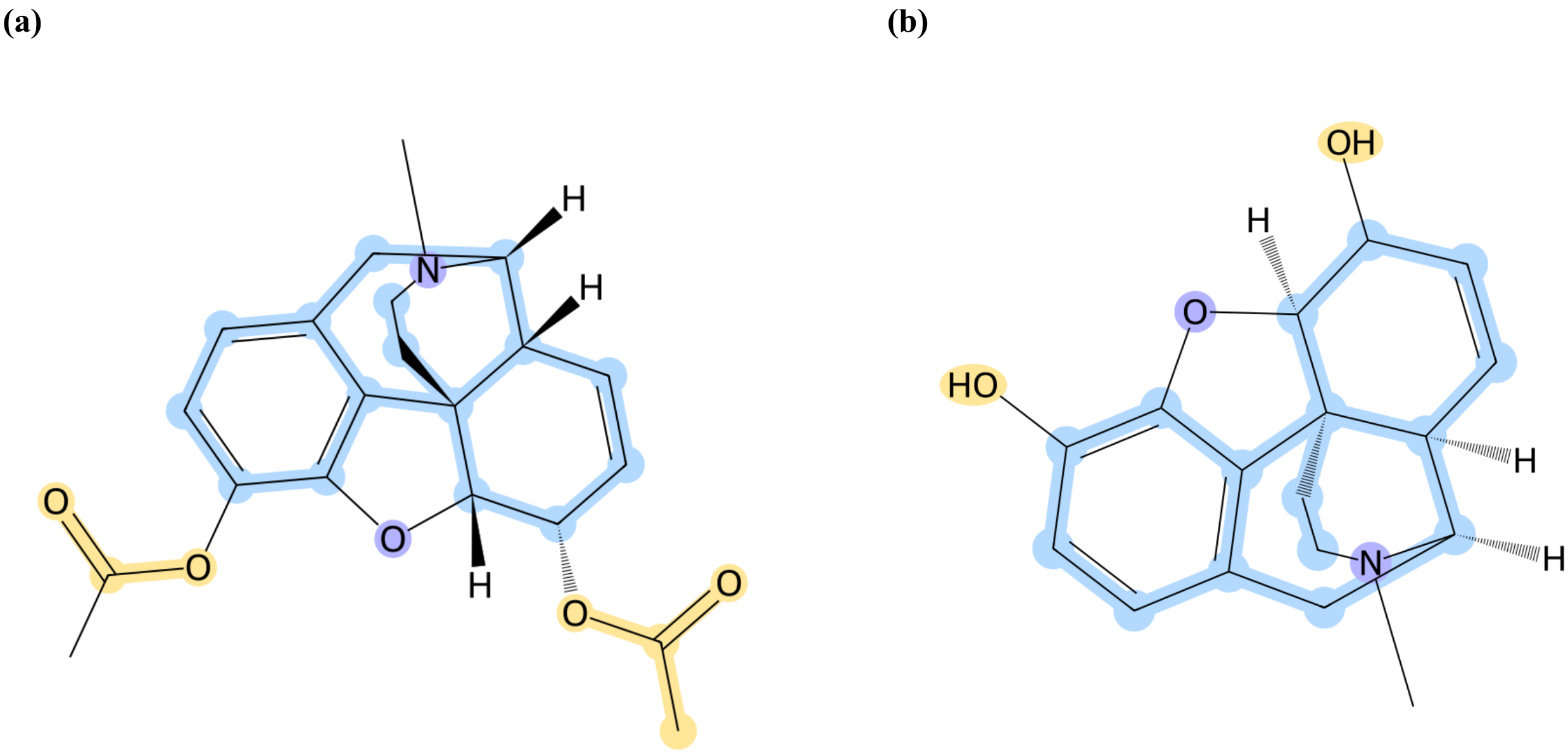

**Figure 6. (a) Heroin structure. (b) Morphine structure.** The blue-highlighted regions mainly covered the shared fused-ring scaffold, indicating that the core morphinan structure contributes substantially to the model prediction, and therefore classified into Group 1: hydrophobic, ring substructures. The yellow-highlighted regions were localized around the substituent groups, suggesting that these functional groups provided additional discriminative information between the two molecules; these were classified into Group 2: oxygen-containing polar substructures. Meanwhile, the purple-highlighted heterocycles were classified into Group 4: Topology- and linker-related patterns.

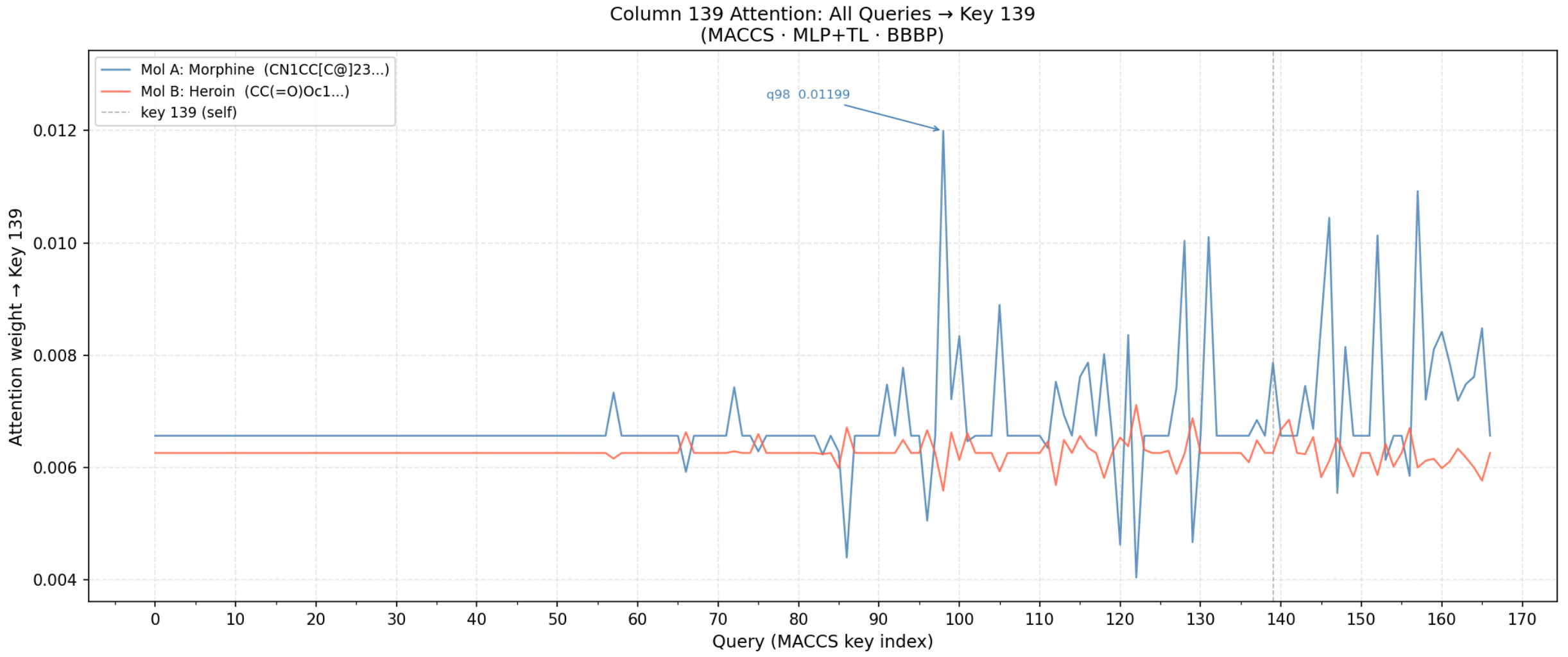

**Figure 7. MACCS attention weights of queries (row) assigned to key 139 (column 139, OH group) for Heroin and Morphine.** The plot illustrated the distribution of attention weights assigned from all query positions to key 139. Compared with Heroin, Morphine showed a more heterogeneous pattern with more pronounced peaks, indicating stronger and more variable attention to the hydroxyl-related feature.

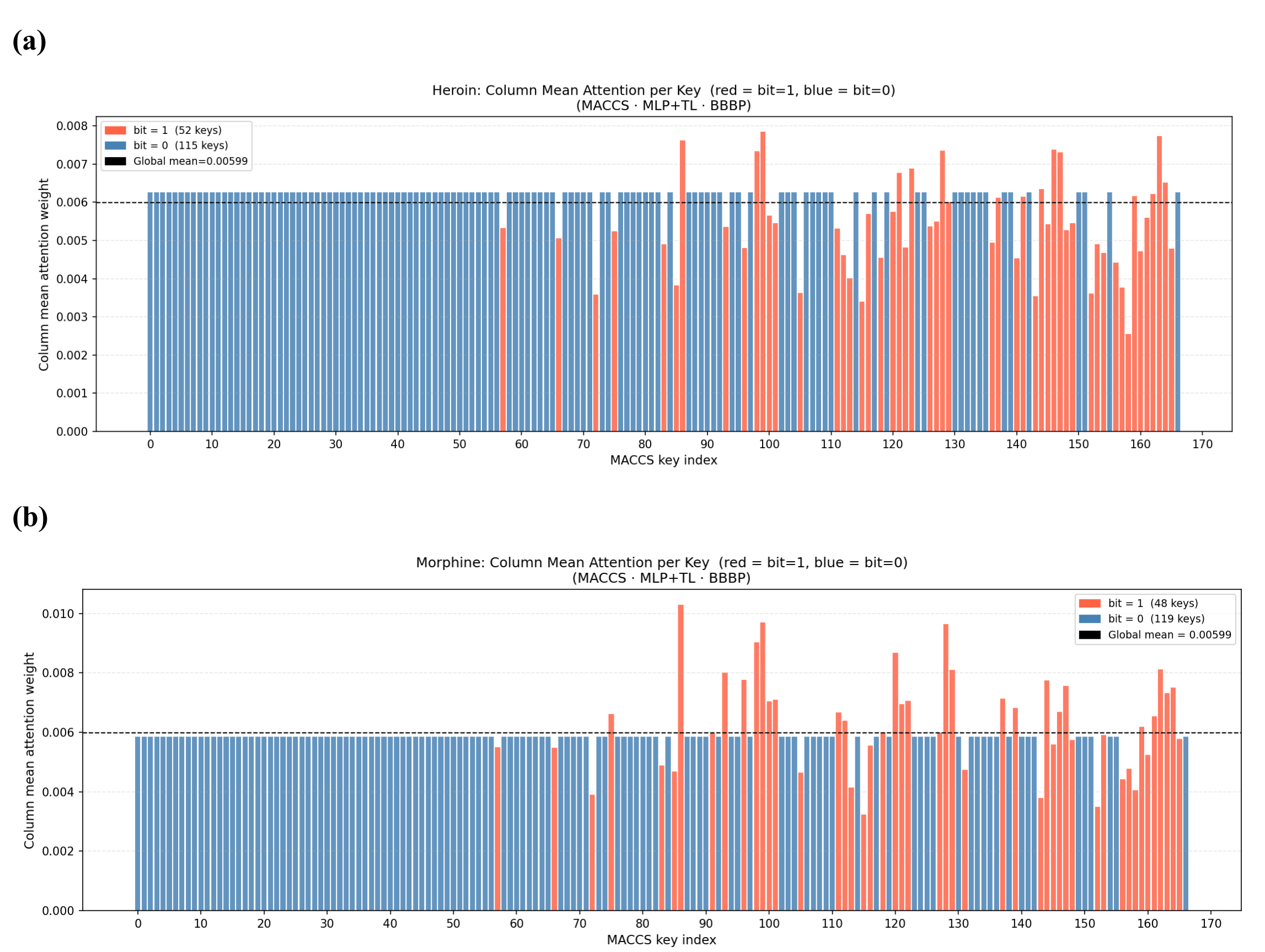

**Figure 8. Column mean attention weights across all 167 MACCS keys for (a) Morphine and (b) Heroin.** The x-axis shows the MACCS key index, and the y-axis shows the column mean attention weight for each key. These plots illustrate that attention was distributed across all keys rather than being restricted only to keys encoded as 1.

# Discussion

In summary, our study should be viewed primarily as a fundamental investigation into the role of molecular encoding in molecular property prediction, rather than as an effort to pursue benchmark-leading performance through increasingly complex model design. While recent advances in molecular property prediction have often emphasized architectural novelty on standard benchmarks, our work instead focused on a more basic but essential question: how different molecular encoding methods influenced prediction, task suitability, and model interpretability across datasets. By systematically comparing multiple encoding methods under two representative neural network-based architectures, we provided evidence that the effectiveness of a fingerprint was not universal, but depended on the target property, dataset characteristics, and modeling framework. In this sense, the contribution of this study lied not only in reporting strong predictive performance, but also in clarifying a foundational issue that was often overlooked in benchmark-driven research. Furthermore, by examining internal attention weights in the Transformer encoder-based framework and relating them to chemically meaningful substructures, we showed that fingerprint studies can also serve as a bridge between predictive modeling and mechanistic interpretation. We therefore hoped that this work can contribute to a more principled understanding of molecular informatics, in which model development was guided not only by performance gains, but also by careful consideration of representation choice, task validity, interpretability, and practical relevance for drug discovery.

Although MLP+TL generally was underperformed compared with the baseline MLP model, it still strongly provided complementary value as an interpretable framework. Particularly, both models (MLP and MLP+TL) performed well on several biologically relevant classification tasks, including toxicity prediction, mutagenicity, and adverse reaction prediction, with average AUC values consistently above 0.9 (Table 4). Interpretability analysis further suggested that Transformer encoder layers could identify dataset-specific important substructures, such as lipophilicity-related features in BBB prediction and nitro-related features in mutagenicity prediction. Taken together, these findings indicated that careful selection of molecular encoding methods and model architecture was more important than simply adopting increasingly complex models, while Transformer-based approaches still offered additional biological insight through substructure-level interpretation.

Comparison with conventional machine learning methods, including XGBoost [58] and Random Forest [65] (Tables S16-S22), showed that traditional ML approaches occasionally performed slightly better predictive performance than neural network-based architectures (MLP and MLP+TL), although the margin was generally small. This observation was consistent with previous large-scale benchmarking studies showing that traditional machine learning models built on fixed molecular representations often remain highly competitive in molecular property prediction, particularly when informative handcrafted descriptors or fingerprints are already provided as inputs [2, 7, 8]. In such settings, the representational burden is partially handled by the fingerprint itself, allowing models such as Random Forest [65] or XGBoost [58] to perform strongly without requiring more complex architectures. From this perspective, the slightly lower performance of our models does not necessarily indicate a limitation of the proposed framework itself but may instead reflect the characteristics of the benchmark setting. When dataset size is limited or the predictive signal is already effectively captured by predefined fingerprints, the additional modeling flexibility offered by neural network-based architectures may not always translate into clear improvements in conventional performance metrics. Prior work has likewise suggested that more complex representation learning models often require sufficient data scale and careful evaluation to demonstrate robust advantages, and that their gains can otherwise be modest or unstable [2, 7]. Importantly, our goal was not solely to maximize predictive performance, but also to investigate whether a neural network-based framework could provide interpretable insights into fingerprint-based prediction. In this regard, the present results align with recent studies emphasizing that model selection in molecular prediction should consider interpretability in addition to predictivity. Although conventional ML models may remain preferable in some tasks when performance alone was prioritized, neural network-based models (MLP and MLP+TL) offered an intermediate alternative by retaining competitive performance while enabling the analysis of internal attention weights to highlight potentially important features.

Several directions may further improve the applicability and generalizability of this framework. First, a limitation of the present study is that model evaluation was performed using five-fold cross-validation under a fixed random seed. Although this setup provides a standard estimate of predictive performance, it may not fully capture the variability introduced by data partitioning. Previous work has shown that different split schemes and random seeds can alter the structural similarity and label divergence among training, validation, and test sets, thereby contributing to performance variance; without sufficient repeated evaluation, apparent metric improvements may partly reflect statistical noise rather than robust methodological gains [7, 66]. In addition, while scaffold-based splitting is widely used as a proxy for chemical-space generalization, it may not fully reflect real-world generalization behavior. The scaffold split mainly probes inter-scaffold generalization, whereas important challenges also arise from intra-scaffold variation, such as activity cliffs, where structurally similar molecules or even molecules sharing the same scaffold can exhibit substantially different properties. Conversely, molecules with different scaffolds may still occupy nearby regions in chemical space. Therefore, scaffold identity alone may not be sufficient to characterize the true difficulty of generalization. More recently, similarity-aware data splitting studies have further emphasized that evaluation can be overly optimistic when highly similar compounds are distributed across training and test sets, because the resulting test performance may partly rely on similarity-induced shortcuts rather than robust out-of-distribution generalization [66]. From this perspective, future work should consider more rigorous evaluation protocols, such as repeated splitting with multiple random seeds, together with reporting mean performance and standard deviation, and potentially incorporating similarity-aware or leakage-reduced splitting strategies. Such designs would provide a more reliable assessment of model stability, uncertainty, and practical generalizability.

Second, future studies may consider selecting between canonical SMILES and isomeric SMILES according to both dataset characteristics and target properties. Canonical SMILES provide a standardized string representation and are therefore commonly adopted in molecular learning tasks, whereas isomeric SMILES preserve stereochemical information that is important for chirality-sensitive endpoints. This issue is especially relevant for BBBP, since previous studies have suggested that BBB permeability can be stereoselective [44]. Nevertheless, the utility of isomeric SMILES may depend on whether the downstream encoding method is capable of preserving stereochemical distinctions. For encoding methods that are largely insensitive to stereochemistry, canonical and isomeric SMILES may yield identical or highly similar feature representations, whereas chirality-aware fingerprints such as stereochemistry-enabled ECFP/FCFP [21] may better capture these differences. This consideration is also important for SMARTS tokenizaion, where preserving stereochemical information require redesigning the tokenization scheme so that stereochemical symbols are explicitly retained.

Third, the current framework inherited an intrinsic dependency from FN [35, 56, 57], namely that auxiliary labels were required together with the primary task label. While this design may enhance prediction by incorporating task-relevant auxiliary information, it also limited the portability of the model to independent datasets in which such annotations were unavailable. This issue is particularly important for external validation, because the practical value of a predictive framework depends not only on its internal performance, but also on whether it can be transferred to new datasets without extensive additional labeling efforts. In principle, this limitation can be addressed in two ways. The first is to obtain ground-truth annotations for the missing auxiliary tasks, which would maintain the fidelity of the original framework but may be labor-intensive and unrealistic in many cases. The second is to infer the missing auxiliary labels using pretrained XGBoost [58] models prior to downstream prediction. This latter strategy is more feasible for large-scale application, yet it transforms the overall framework into a two-stage pipeline in which uncertainty from the auxiliary-label prediction step may propagate to the final endpoint. Consequently, the benefits of auxiliary information must be weighed against the risk of accumulated prediction error. Future studies should therefore move beyond reporting overall accuracy and examine this dependency more explicitly. For example, it would be informative to quantify how sensitive the final model is to noise or bias in the imputed auxiliary labels, whether different auxiliary tasks contribute unequally to performance gains, and whether model interpretability remains stable when auxiliary information is inferred rather than experimentally observed. Addressing these questions would provide a clearer understanding of the boundary between methodological effectiveness and practical applicability and would help determine how FN can be adapted for more realistic external prediction scenarios. In short, our study highlights molecular encoding as a fundamental issue in molecular property prediction, and future work should focus on more rigorous evaluation protocols, stereochemistry-aware representations for chirality-sensitive endpoints, and a deeper understanding of how auxiliary-label dependency shapes the robustness, interpretability, and practical generalizability of FN.